\documentclass[10pt,journal,compsoc]{IEEEtran}
\ifCLASSOPTIONcompsoc
  \usepackage[nocompress]{cite}
\else
  \usepackage{cite}
\fi

\usepackage{subfigure}
\usepackage{graphicx}
\usepackage{amsmath}
\usepackage{amssymb}
\usepackage{hyperref}
\usepackage{booktabs}
\usepackage{stfloats}
\usepackage{dsfont}
\usepackage{graphicx}
\usepackage{subcaption}
\usepackage{xcolor}

\begin{document}

\title{EdgeLoc: A Communication-Adaptive Parallel System for Real-Time Localization in Infrastructure-Assisted Autonomous Driving}

\author{Boyi Liu,
        Jingwen Tong,
        Yufan Zhuang
\IEEEcompsocitemizethanks{\IEEEcompsocthanksitem The authors information. \protect\\

% note need leading \protect in front of \\ to get a newline within \thanks as
% \\ is fragile and will error, could use \hfil\break instead.
% E-mail: see http://www.michaelshell.org/contact.html
}% <-this % stops a space
% \thanks{Manuscript received April 19, 2005; revised August 26, 2015.}
}

% The paper headers
% \markboth{Journal of \LaTeX\ Class Files,~Vol.~14, No.~8, August~2015}%
% {Shell \MakeLowercase{\textit{et al.}}: Bare Advanced Demo of IEEEtran.cls for IEEE Computer Society Journals}

\IEEEtitleabstractindextext{%
\begin{abstract}
This paper presents EdgeLoc, an infrastructure-assisted, real-time localization system for autonomous driving that addresses the incompatibility between traditional localization methods and deep learning approaches. The system is built on top of the Robot Operating System (ROS) and combines the real-time performance of traditional methods with the high accuracy of deep learning approaches. The system leverages edge computing capabilities of roadside units (RSUs) for precise localization to enhance on-vehicle localization that is based on the real-time visual odometry. EdgeLoc is a parallel processing system, utilizing a proposed uncertainty-aware pose fusion solution. It achieves communication adaptivity through online learning and addresses fluctuations via window-based detection. Moreover, it achieves optimal latency and maximum improvement by utilizing auto-splitting vehicle-infrastructure collaborative inference, as well as online distribution learning for decision-making. Even with the most basic end-to-end deep neural network for localization estimation, EdgeLoc realizes a 67.75\% reduction in the localization error for real-time local visual odometry, a 29.95\% reduction for non-real-time collaborative inference, and a 30.26\% reduction compared to Kalman filtering. Finally, accuracy-to-latency conversion was experimentally validated, and an overall experiment was conducted on a practical cellular network. The system is open sourced at \href{https://github.com/LoganCome/EdgeAssistedLocalization}{https://github.com/LoganCome/EdgeAssistedLocalization}.
\end{abstract}

}
\maketitle

\IEEEdisplaynontitleabstractindextext

\IEEEpeerreviewmaketitle

\ifCLASSOPTIONcompsoc
\IEEEraisesectionheading{\section{Introduction}\label{sec:introduction}}
\else
\section{Introduction}
\label{sec:introduction}
\fi

\IEEEPARstart{L}{ocalization} is pivotal for autonomous vehicles, laying the groundwork for subsequent navigational tasks. Traditional systems, employing sensor fusion via Kalman filters, combine data from Inertial Measurement Units (IMU), Global Positioning System (GPS), and visual odometry (VO) for consistently accurate positioning. Recently, deep neural networks (DNNs) have emerged, offering end-to-end (E2E) vision-based localization solutions capable of directly computing the vehicle's absolute pose with remarkable precision. Nevertheless, their applications within autonomous driving are hindered by concerns over reliability, computational demands, and latency during inference.

\begin{figure}
  \centering
  \includegraphics[width=0.5\textwidth]{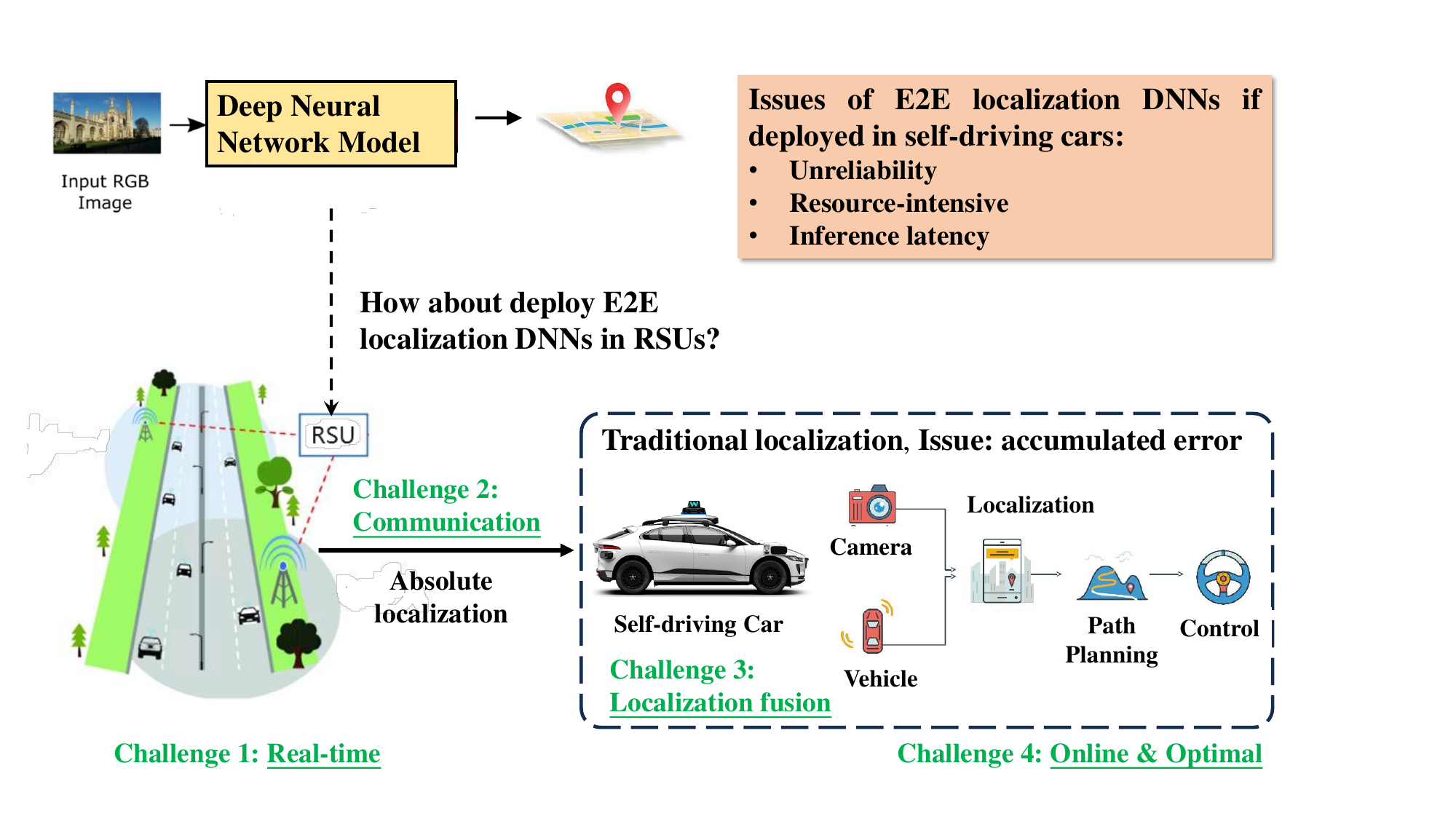}
  \caption{Challenges and system overview of deploying end-to-end localization deep neural networks on roadside units for infrastructure-assisted autonomous driving.}
  \label{fig:introduction}
\end{figure}

As shown in Fig. \ref{fig:introduction}, our paper bridges the gap between traditional and deep learning localization methods by proposing the deployment of E2E localization DNNs on roadside units (RSUs), leveraging their edge computing capabilities to bolster autonomous driving localization. This edge-assisted innovative approach, however, introduces a set of challenges: \textbf{ 1) ensuring real-time operations, 2) establishing robust vehicle-to-infrastructure communications, 3) integrating RSU-provided localization with on-vehicle systems, and 4) optimizing the entire framework for live use without sacrificing accuracy or safety}.

To address these challenges, we develop EdgeLoc, a communication-adaptive parallel system for real-time localization in infrastructure-assisted autonomous driving. Unlike traditional infrastructure-assisted methods focusing on minimizing latency for real-time performance but suffering from uncontrollable network conditions, we propose a novel ROS-based system framework that incorporates two parallel threads: real-time visual odometry and non-real-time DNN inference. By integrating the real-time performance of traditional methods with the edge computing-enhanced high accuracy of deep learning approaches, our system leverages precise localization from a roadside unit as the infrastructure to improve localization based on the real-time on-vehicle visual odometry. This provides a novel and practical solution for utilizing DNN models in real-time localization for autonomous driving.

The implementation of EdgeLoc encompasses several key developments:

\begin{itemize}
    \item The design and execution of a parallel processing system harmonizes VO and DNN inference, optimizing both for real-time performance and accuracy.
    \item The development of the pose fusion technique circumvents the limitations of Kalman filtering with DNNs, addressing the challenges posed by the black-box nature of DNNs and the potential for outliers in pose outputs.
    \item It is critical to enhance system robustness to withstand variations in communication status, enabling a crucial factor for the unpredictability of real-world vehicle-to-infrastructure (V2I) communication scenarios.
    \item Ensuring that the framework not only improves on-vehicle localization accuracy but also maximizes this enhancement under operational conditions.
\end{itemize}

In this work, we address the above key technologies and implement EdgeLoc. The main contributions are as follows:
\begin{enumerate}
    \item We propose a ROS-based framework comprising RSU, VO, and decision modules to enable parallel collaborative computation under an action mechanism.
    \item A novel method is proposed to estimate the uncertainty of poses from the RSU, addressing the reliability concerns associated with the black-box nature of deep learning models and potential outliers in pose outputs. Notably, the uncertainty estimation approach recognizes that localization accuracy is contingent on both RSU latency and DNN pose uncertainty estimation. Therefore, the estimation method is uniquely correlated to latency, thereby transforming the localization accuracy optimization into a single objective of inference latency optimization.
    \item Through auto-splitting of vehicle-infrastructure collaborative inference, the system achieves optimal latency, and the maximum localization improvement is guaranteed via online learning for decision-making.
    \item The online learning module seamlessly handles communication state transitions and incorporates a window-based out-of-distribution solution to fortify robustness against network fluctuations.
\end{enumerate}

The remainder of this paper is organized as follows: Section \ref{section2} presents the related work in this field. Section \ref{section3} describes how EdgeLoc achieves real-time performance and solves Challenge 1 shown in Fig. \ref{fig:introduction}. Section \ref{section4} explains how EdgeLoc performs localization fusion and solves Challenge 3 shown in Fig. \ref{fig:introduction}. Section \ref{section5} discusses how EdgeLoc optimizes communication decision-making and solves Challenge 2 shown in Fig. \ref{fig:introduction}. Section \ref{section7} concludes the paper.
% the conclusion. 
Appendix \ref{appendix} provides proof of the optimality of the proposed method, addressing challenge 4 shown in Fig. \ref{fig:introduction}.
\section{Related Works}\label{section2}
\subsection{Localization in Autonomous Driving}
Localization is a crucial task in autonomous driving \cite{woo2018localization}, ensuring that vehicles ascertain their exact localization within their surroundings. Existing methods can be categorized into sensor-based integral localization methods, such as visual odometry and LiDAR-based techniques, and absolute localization systems like GPS. Recently, DNN-based methods have emerged as a promising approach for end-to-end localization in modern autonomous vehicles.
\subsubsection{Sensor-based Integral Localization}\label{section_2_1_1}
Relative localization techniques employ sensors like IMUs, cameras, and LiDARs for real-time position estimation. For example, IMUs measure acceleration and angular velocity via accelerometers and gyroscopes, respectively, and integrate these data to estimate the vehicle's position, velocity, and orientation over time. However, this integration process can lead to drift and error accumulation due to noise and biases. 
VO \cite{fraundorfer2012visual} analyzes camera image sequences to compute motion by extracting and matching visual features between consecutive frames, and then integrates these estimates to track the vehicle's trajectory \cite{qin2018vins, zhan2020visual}. 
Despite its higher accuracy, VO is prone to cumulative errors in feature-scarce environments \cite{fraundorfer2012visual}. 
On the other hand, LiDAR generates detailed 3D maps through laser scanning, producing point clouds that aid in accurate trajectory mapping \cite{shan2020lio}. Although effective in diverse conditions, LiDAR demands significant computational resources.

Given the limitations of individual sensors regarding error accumulation and environmental dependency \cite{agostinho2022practical,de2020evaluating}, sensor fusion techniques such as Kalman filters or particle filters are commonly employed. These algorithms integrate data from multiple sensors like IMUs, VO, LiDAR, and radar, to produce more reliable and accurate localization estimates, thereby overcoming the shortcomings of single-sensor systems.

%each sensor type.

\subsubsection{Absolute Localization}
Absolute localization techniques determine geographic coordinates using external data such as satellite systems or pre-mapped environments. Two primary approaches are Global Navigation Satellite Systems (GNSS) and DNN-based methods \cite{kendall2015posenet}.

The most widely used GNSS is GPS, which calculates position through trilateration using signals from multiple satellites \cite{zhao2021gnss}. In addition, the robustness and accuracy of the GPS localization can be enhanced by integrating GPS with inertial navigation systems (INS) \cite{jiang2022ins}. The INS compensates for GPS signal loss, while GPS bound the error accumulation in INS.

DNN-based end-to-end localization directly regresses camera pose from images. PoseNet \cite{kendall2015posenet} pioneered this, inspiring further works to improve performance through geometric loss functions \cite{kendall2017geometric}, uncertainty modeling \cite{kendall2017uncertainties}, and LSTM units \cite{walch2017image}. Enhancements in network architectures like hourglass networks \cite{melekhov2017image} and MapNet \cite{brahmbhatt2018geometry} have further improved localization performance.

Nonetheless, these absolute localization methods encounter significant challenges. The GPS accuracy can be affected by factors such as signal blockage, multipath effects, and atmospheric conditions, particularly in urban environments with tall buildings or tunnels \cite{lu2022gnss}. On the other hand, DNN-based localization methods face challenges in terms of computational efficiency, latency, and robustness to adverse weather conditions. Therefore, it is necessary to develop multiple-sensor fusion or infrastructure-assisted approaches to address these issues.

\subsection{Infrastructure-assisted Autonomous Driving}
Infrastructure-assisted autonomous driving represents a promising paradigm to enhance the safety, efficiency, and reliability of autonomous vehicles. Utilizing roadside infrastructure, such as sensors, communication devices, and computing units, this approach not only provides additional contextual information about the vehicle's surroundings but also significantly improves its localization precision \cite{cress2023intelligent}. 
By seamlessly integrating data from infrastructure-based sensors with onboard systems, autonomous vehicles can achieve more accurate positioning. This integration extends their sensing range and enhances their perception capabilities \cite{liu2021peer}. The collaborative model improves situational awareness \cite{liu2019federated, he2021vi} and facilitates coordinated decision-making between vehicles and infrastructure \cite{liu2022communication}. As a result, this leads to safer, more efficient computing \cite{liu2023roboec2} and better road navigation \cite{callegaro2020dynamic, liu2019lifelong}.

Recent research has explored the potential of infrastructure-assisted autonomous driving in various contexts. For instance, Shi et al. \cite{shi2022vips} developed VIPS, a real-time perception fusion system that achieves decimeter-level accuracy and low latency, demonstrating the feasibility of integrating infrastructure-based perception with vehicle-based systems. Liu et al. \cite{liu2022communication} proposed a framework for infrastructure-assisted autonomous vehicles, addressing challenges in bandwidth and latency variations.
Their work emphasizes the importance of optimizing communication protocols and architectures. Moreover, Yang et al. \cite{yang2022scalable} studied infrastructure-vehicle collaborative autonomous driving, considering scenarios with a large number of RSUs and vehicles, providing insights into the design and deployment of large-scale V2X networks. Jia et al. \cite{jia2024infrastructure} demonstrated the benefits of infrastructure-assisted collaborative perception in automated valet parking, improving safety and increasing the maximum safe cruising speed. Tsukada et al. \cite{tsukada2020networked} introduced networked roadside perception units (RSPUs) for infrastructure-based cooperative perception, demonstrating advanced performance in delivering messages to autonomous vehicles within strict latency constraints. In addition, Hirata et al. \cite{hirata2021roadside} proposed a cooperative path-planning model for autonomous vehicles at intersections using future path sharing based on information from RSUs, showing improved safety and efficiency. These studies underscore the significance of real-time data processing, communication, and cooperative planning in enhancing the capabilities and performance of autonomous vehicles in complex environments.

Despite the advancements in infrastructure-assisted localization, there are still limitations, such as high infrastructure costs, the need for optimal configuration of V2I networks, and the lack of effective fusion between infrastructure-based and vehicle-based techniques. To address these limitations, the proposed EdgeLoc system aims to integrate infrastructure-assisted and vehicle-based techniques using a parallel processing architecture. By combining the benefits of high-precision absolute localization from RSUs with real-time relative localization from onboard visual odometry, EdgeLoc achieves centimeter-level accuracy at a lower cost while exhibiting strong robustness and adaptability to dynamic environments.
\section{Real-Time Parallel Architecture of EdgeLoc}\label{section3}

\begin{figure*}[ht]
  \centering
  \includegraphics[width=1\textwidth]{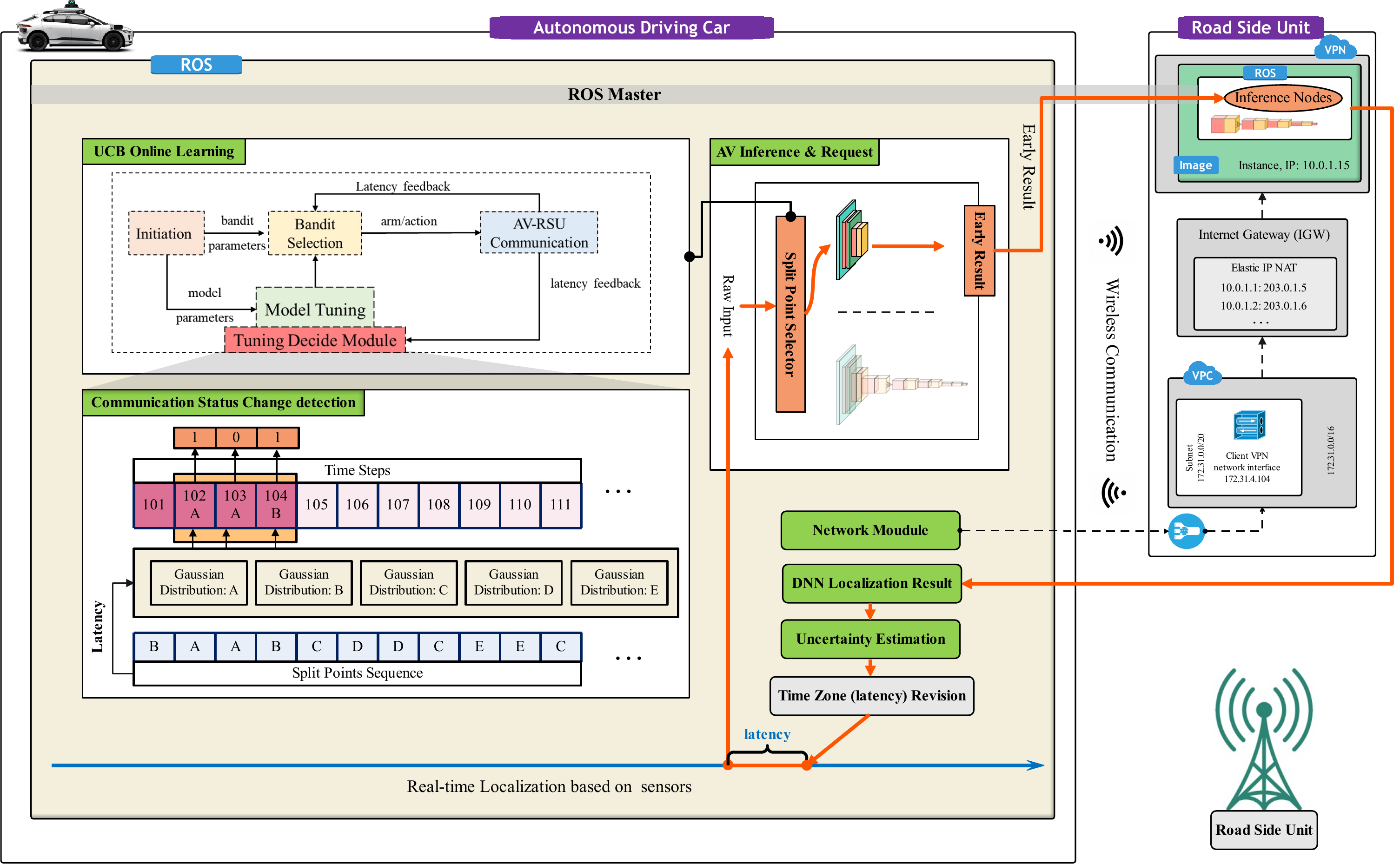}
  \caption{System architecture of EdgeLoc: the core components, communication setup, and collaborative operations between the autonomous vehicle and roadside unit.}
  \label{fig:system}
\end{figure*}

The real-time performance of the EdgeLoc system is achieved through a novel system architecture and workflow optimization. Our system employs a parallel processing architecture that is built on top of ROS to seamlessly integrate real-time localization from on-vehicle sensors with precise localization enhancements from roadside infrastructure. By leveraging ROS's ability to efficiently manage parallel processes and communication between distributed nodes, our system achieves high performance and low latency. As illustrated by the red arrows in Fig. \ref{fig:system}, the autonomous vehicle (AV) performs real-time localization based on its onboard sensors, while simultaneously interacting with the RSU for collaborative inference to improve accuracy.
\subsection{System Architecture and Core Modules}
The architecture of the EdgeLoc system, depicted in Fig. \ref{fig:system}, is designed to augment real-time localization capabilities for AVs. This section presents the key modules that ensure the system's robust and adaptive performance, even in dynamic network environments.

\textbf{System overview:} The operational paradigm of the EdgeLoc system is as follows: The AV conducts real-time localization through its onboard sensors and the preliminary layers of the DNN designed for localization. The AV Inference \& Request module autonomously partitions the DNN model at an optimal split point determined by the UCB online learning module. The initial inference results are transmitted from the AV to the RSU via the ROS Master communication protocol. Subsequently, the RSU, equipped with sufficient computing resources, processes the remaining layers of DNN inference and sends the localization results back to the AV. 
The UCB module continuously learns and adjusts its decision-making model for DNN partitioning in response to the prevailing network conditions and observed latencies. A communication status change detection submodule employs a sliding window technique to monitor latency distributions and activates the UCB module to refine its decision model upon detecting substantial shifts in network conditions. In scenarios where AVs and RSUs operate across disparate network segments, Virtual Private Network (VPN) and Network Address Translation (NAT) mechanisms are employed to facilitate uninterrupted communication by converting private IP addresses into public IP addresses.

The \textbf{AV Inference \& Request Module} plays a pivotal role in the system by dynamically partitioning the DNN localization model. This module intelligently adapts the computational workload distribution between the AV and the RSU based on the prevailing network conditions and available resources. By transmitting early inference results to the RSU via the ROS Master communication mechanism, EdgeLoc effectively minimizes latency while maximizing localization accuracy.

The \textbf{UCB Online Learning Module} is built upon the principles of reinforcement learning and multi-armed bandit algorithms. The UCB algorithm \cite{zhou2015survey} is a well-established approach for balancing exploration and exploitation in decision-making processes. In the context of EdgeLoc, the UCB module learns a decision model that maps the current network status to the optimal split point for the DNN model. The algorithm maintains a confidence bound for each potential split point and updates these bounds based on the observed latencies. By selecting the split point with the highest UCB, the UCB module ensures that the system explores new split points while exploiting the knowledge gained from previous observations. The UCB algorithm has been theoretically proven to achieve logarithmic regret, guaranteeing near-optimal performance over time.

The \textbf{Communication Status Change Detection Module} employs advanced statistical techniques to detect significant changes in network conditions. The submodule maintains Gaussian distributions of the observed latencies for each split point and utilizes the Kullback-Leibler (KL) divergence to measure the difference between the distributions. The KL divergence is a well-established metric in information theory that quantifies the dissimilarity between two probability distributions. By comparing the KL divergence between the current and previous latency distributions, the submodule can detect significant shifts in network conditions. To mitigate the impact of transient fluctuations, the submodule employs a sliding window technique, requiring consistent distribution differences over multiple consecutive frames before triggering a change detection. This approach ensures that the system adapts to genuine changes in network conditions while filtering out temporary disturbances.

To address the temporal misalignment between the AV and RSU localization results, EdgeLoc incorporates a \textbf{Time Zone Localization Result Revision Module}. This module employs a two-stage correction process to align the localization results obtained from the RSU with the current time step of the AV. The mathematical formulas governing this correction process are elegantly illustrated in the light red background of Fig. \ref{fig:system}.

On the RSU side, the \textbf{RSU Inference Module} is responsible for completing the remaining layers of the DNN inference upon receiving the early inference results from the AV. This module leverages the computational resources of the RSU to perform the final stage of the E2E localization, generating the DNN based localization outputs that are subsequently transmitted back to the AVs.

In scenarios where the AVs and RSUs are not within the same network, EdgeLoc seamlessly integrates network address translation mechanisms to facilitate communication. By establishing a VPN, EdgeLoc enables seamless interaction between devices in different subnets, assigning virtual IP addresses to allow communication as if they were in the same local network. The system utilizes an internet gateway and Elastic IP NAT to map the private IP addresses of the AVs and RSUs to public IP addresses, enabling communication across different network segments.

By leveraging this processing architecture and the collaborative inference approach, EdgeLoc achieves real-time, high-accuracy localization for autonomous vehicles. The system intelligently adapts to network conditions, minimizes latency, and ensures robust performance even in the presence of network fluctuations.

\subsection{Parallelism of EdgeLoc}

\begin{figure}
  \centering
  \includegraphics[width=0.5\textwidth]{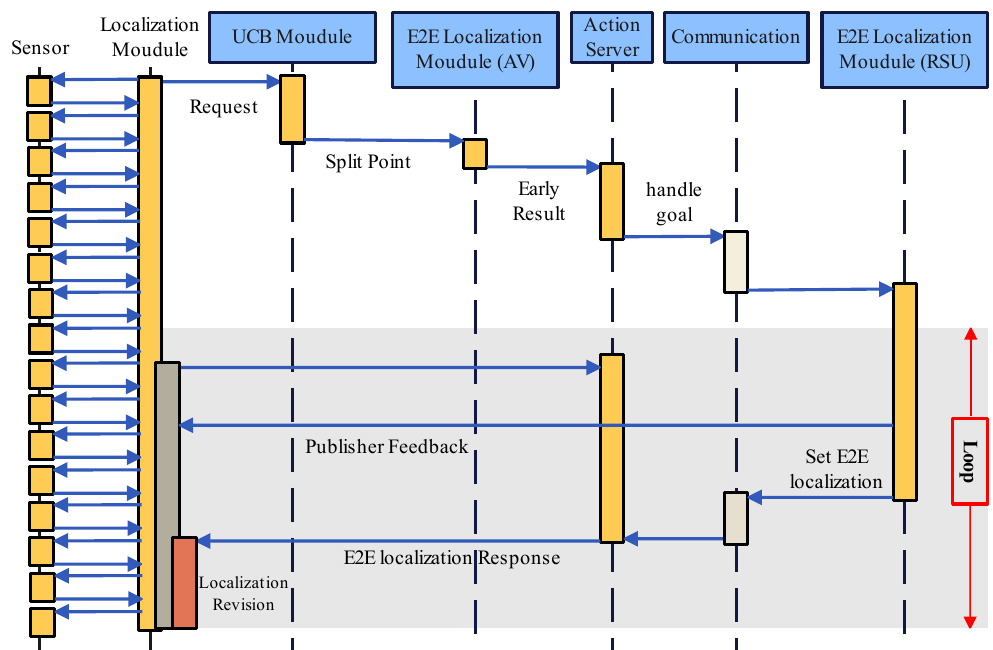}
  \caption{The diagram includes five modules (end-to-end localization modules on the vehicle and RSU, sensor loop module, localization revision module, and action server), vertical lines representing the lifecycle of each module, arrows showing interactions and invocations between modules, and the parallel execution of real-time localization and RSU collaboration threads.}
  \label{fig:thread}
\end{figure}

EdgeLoc is designed with a focus on parallelism to achieve real-time performance. The system architecture, as illustrated in the sequence diagram (Fig. \ref{fig:thread}), demonstrates two parallel threads that work in tandem to provide efficient and accurate localization results.

The first thread, based on the vehicle's onboard sensors, operates in a continuous loop to handle the real-time localization tasks. This thread is responsible for processing the data from various sensors such as camera, IMU, or LiDAR, to estimate the vehicle's position and orientation. The localization module in AV performs the necessary computations and updates the localization results in real-time through integral calculation as described in Section \ref{section_2_1_1}. This thread ensures that the AV always has access to the most recent localization information, enabling it to receive real-time localization results.

The second thread, involving the communication between the AV and the RSU, runs in parallel to the sensor-based localization thread. This thread is triggered when the AV sends a request to the UCB module for determining the optimal splitting point of the DNN model. Based on the decision from the UCB module, the AV performs the initial inference up to the determined split point and sends the intermediate results to the RSU. The RSU then continues the DNN inference using its computational resources, and sends the DNN's localization result back to the AV. The AV's localization module integrates this high-precision information with its own sensor-based estimate to obtain a refined and more accurate positioning result, benefiting from the collaborative inference process without requiring powerful onboard computing hardware.

The RSU's localization module leverages the DNN model to process the data received from the AV and provide an DNN based localization result. The DNN takes the AV's sensor data as input and outputs a refined pose estimate by extracting relevant features and learning the complex mapping between the input data and the vehicle's position. This approach allows the RSU to utilize its computational resources to perform the computationally intensive DNN inference and a more accurate localization result. By offloading some of the DNN computation to the RSU, the AV can benefit from the improved accuracy without the need for powerful onboard computing hardware, making the system more cost-effective and efficient. This parallel thread enables the AV to leverage the RSU's processing capabilities and the DNN's ability to learn from data, ultimately enhancing the overall localization performance.

Once the RSU's localization module completes its processing, it sends the localization result back to the AV. The AV's localization module receives this result and integrates it with its own sensor-based localization estimate. This fusion of the two localization results is performed using our proposed method in Section \ref{section4}, to obtain a refined and more precise localization estimate. The updated localization information is then published to other modules in the AV, such as the perception, planning, and control modules, to facilitate accurate and safe autonomous driving.

The parallelism of EdgeLoc offers several advantages. First, it allows the AV to maintain real-time localization capabilities even in challenging environments where sensor-based localization alone may be insufficient. Second, it enables the AV to leverage the computational resources and additional information available at the RSU, reducing the burden on the AV's onboard systems. Third, the parallel architecture ensures that the localization process remains resilient and fault-tolerant. If one thread experiences issues or delays, the other thread can continue to provide localization results, maintaining the overall system's robustness.

Moreover, the parallel design of EdgeLoc facilitates scalability and adaptability. As the autonomous driving infrastructure evolves and expands, additional RSUs can be deployed to cover larger areas and provide localization assistance to a growing fleet of AVs. The parallel architecture allows for seamless integration of new RSUs and the efficient distribution of localization tasks across multiple units.

In short, the parallelism of EdgeLoc is a key aspect of its design, enabling real-time localization in infrastructure-assisted autonomous driving. The two parallel threads, one based on the AV's sensors and the other involving AV-RSU communication, work together to provide accurate, reliable, and timely localization results. This parallel architecture enhances the system's accuracy, robustness, scalability, and adaptability, making EdgeLoc a promising solution for the localization challenges in autonomous driving.
\section{Localization Fusion of EdgeLoc}\label{section4}

In this section, we introduce the localization fusion algorithm implemented in EdgeLoc. 
The unsuitability of the Kalman filter motivates us to propose a method compatible with EdgeLoc, which transforms the localization accuracy problem into a latency reduction problem.

\subsection{Limitations of Kalman Filtering in End-to-End Localization}
Kalman filtering is a widely used sensor data fusion technique in traditional localization systems. However, significant challenges arise when applying it to localization based on DNNs. The primary obstacle is the difficulty in estimating the Kalman gain of DNNs, which is crucial for the proper functioning of the Kalman filter.

In Kalman filtering \cite{simon2001kalman}, the estimation of the system state relies on prior knowledge of the process noise covariance \(Q\) and the measurement noise covariance \(R\). These covariances are used to determine the optimal Kalman gain \(K\), which balances the predicted state and the measured state. The update equations for the Kalman filter can be rewritten as follows to match the symbols in the subsequent localization equations:
\begin{equation}
\begin{aligned}
L_{r,k|k-1} &= A L_{r,k-1|k-1} + B L_{\beta,k}, \\
P_{k|k-1} &= A P_{k-1|k-1} A^T + Q, \\
K_k &= P_{k|k-1}(P_{k|k-1} + R)^{-1}, \\
L_{r,k|k} &= L_{r,k|k-1} + K_k(L_{\alpha,k} - L_{r,k|k-1}), \\
P_{k|k} &= (I - K_k)P_{k|k-1}.
\end{aligned}
\end{equation}
In the above formulas, \( L_r \) represents the fused localization result, \( L_\alpha \) represents the DNN-based localization output, \( L_\beta \) represents the real-time localization result from on-vehicle sensors, \( P \) represents the state covariance, \( A \) and \( B \) are system matrices, and \( I \) is the identity matrix, $k$ is the time step.

However, when dealing with DNN-based localization, the measurement noise covariance \(R\) cannot be reliably estimated. DNNs are often regarded as black-box models, making it challenging to characterize their error distribution. Unlike traditional sensors with well-defined noise models, DNN outputs may exhibit complex and unpredictable error patterns. The lack of a reliable noise model impedes the application of Kalman filtering to fuse DNN localization results with other sensor data.

Let \(e_k\) represent the measurement error in DNN localization:
\begin{equation}\label{ek}
e_k = L_{\alpha,k} - L_{r,k|k-1}.
\end{equation}
For traditional sensors, the measurement error typically follows a zero-mean Gaussian distribution, i.e., \(e_k \sim N(0, R)\). However, for DNN localization, due to its black-box nature, the measurement error distribution is unknown and unpredictable. We assume the true error distribution is \(e_k \sim N(\mu_k, \Sigma_k)\), where \(\mu_k\) and \(\Sigma_k\) are the unknown mean and variance of the DNN outputs.

Substituting equation \ref{ek} into the Kalman filtering update equation, we get:
\begin{equation}\label{Rkalm}
\begin{aligned}
L_{r,k|k} &= L_{r,k|k-1} + K_k(L_{\alpha,k} - L_{r,k|k-1}) \\
&= L_{r,k|k-1} + K_ke_k \\
&= L_{r,k|k-1} + K_k(\mu_k + \varepsilon_k),
\end{aligned}
\end{equation}
where \(\varepsilon_k \sim N(0, \Sigma_k)\). From equation \ref{Rkalm}, it is evident that if the DNN localization exhibits an unknown bias \(\mu_k \neq 0\), the Kalman filter estimation will incorporate this bias term \(K_k\mu_k\). This means that the Kalman filter introduces anomalies and biases from DNN localization into the final fused localization result. Additionally, since \(\Sigma_k\) is unknown, the calculation of the Kalman gain \(K_k\) becomes unreliable, further degrading the performance of fused localization.

Due to the recursive nature of the Kalman filter, any error in state estimation not only affects the current state prediction and update but also propagates through the entire system over time. If the system matrices \(A\), or the control inputs \(B\) do not adequately compensate for these errors, or if the error sources are persistent, this can lead to cumulative errors and eventual system instability. Therefore, applying Kalman filtering to fusion with DNN-based localization presents inherent challenges. In DNN localization, the uncertainty, which Kalman filtering treats, cannot be effectively modeled and managed by the filters, leading to reduced performance in fusion localization. Consequently, it is necessary to propose a localization fusion method adapted to EdgeLoc.
\subsection{The Proposed Localization Fusion Method Adapted to EdgeLoc}
To address the limitations of Kalman filtering and enable effective fusion of DNN-based localization results, we propose a novel fusion method tailored to the characteristics of DNN outputs. Our method leverages the temporal consistency of localization results and introduces an uncertainty estimation technique that adapts to the latency of DNN inference.
The proposed fusion method is as follows:
\begin{multline}\label{localization}
% equation --  aligned
\begin{cases}
L_r(t) = u L_\alpha(t) + (1-u) L_r(t-\Delta t), \\ 
\qquad \text{if } t = x \Delta t, x \in \mathbb{N} \\
L_r(t+a) = L_r(t+a-1) + L_\beta(t+a) - L_\beta(t+a-1), \\
\qquad \text{if } t \neq x \Delta t, x \in \mathbb{N}, a \in [1, \Delta t], \\
\end{cases}\nonumber
\end{multline}
where $L_r(t)$ represents the fused localization result at time $t$, $L_\alpha(t)$ represents the DNN-based localization output at time $t$, $L_r(t-\Delta t)$ represents the fused localization result at the previous time step, and $L_\beta(t)$ represents the real-time localization result from on-vehicle sensors at time $t$.
Specifically, the fusion process operates in two distinct cases:
\begin{itemize}
    \item When $t = n \Delta t$, where $n \in \mathbb{N}$, EdgeLoc fuses the DNN-based localization output $L——\alpha(t)$ and the previous result $L_r(t-\Delta t)$. The parameter $u$ is a weighted coefficient determined by the estimated output uncertainty of the DNN.
    % represents the weight assigned to the DNN output, which is determined based on the estimated uncertainty of the DNN result.
    \item When $t \neq n \Delta t$, the fused localization result is updated incrementally using the difference between the real-time localization $L_\beta(t)$ and its corresponding value at the fusion time step $L_\beta(t+i)$, where $i \in [1, \Delta t]$.
\end{itemize}
One key component in our fusion method is the uncertainty estimation of the DNN-based localization output. We propose a latency-aware uncertainty estimation technique that adapts to the inference time of the DNN. The uncertainty estimation $\Lambda$ and the weight $u$ are given by
\begin{equation}
\Lambda = \frac{1}{1 + e^{-k(\Delta t - \Delta t_0)}},
\end{equation}
and
\begin{equation}\label{weight}
u = f(\Delta t) = 1 - \Lambda,
\end{equation}
where $\Delta t$ represents the DNN inference latency, $\Delta t_0$ is a reference latency value, and $k$ is a scaling factor. The function $f(\Delta t)$ is a sigmoid function that maps the latency to an uncertainty value in the range of 0 to 1. As the latency increases, the uncertainty of the DNN output increases and the weight $u$ in \eqref{weight} decreases. 
% To reflect the relationship between $u$ and $L_\alpha(t)$, we have defined Eq. \eqref{weight}.

\subsection{Improving Localization by Transforming Accuracy Improvement to Latency Reduction}

By integrating latency-aware uncertainty estimation into the fusion process, our proposed method effectively reformulates the problem of improving localization accuracy as a latency reduction problem in DNN inference. Rather than directly targeting accuracy improvement, our approach focuses on minimizing the latency of DNN inference, which in turn leads to enhanced overall localization accuracy.

The relationship between latency and accuracy can be derived from the fusion equation:
\begin{equation}
\begin{split}
L_r(t) &= u L_\alpha(t) + (1-u) L_r(t-\Delta t) \\
&= f(\Delta t) L_\alpha(t) + (1-f(\Delta t)) L_r(t-\Delta t) \\
&= \frac{L_r(t-\Delta t)}{1 + e^{-k(\Delta t - \Delta t_0)}}  + \left(1 - \frac{1}{1 + e^{-k(\Delta t - \Delta t_0)}}\right) L_\alpha(t).
\end{split}
\end{equation}
As the latency $\Delta t$ decreases, the weight assigned to the DNN output $L \alpha(t)$ increases, leading to a higher contribution of the DNN result to the fused localization. Conversely, as the latency increases, the weight of the DNN output decreases, and the fused localization relies more on the previous fused result and the real-time localization from on-vehicle sensors.

The latency-accuracy relationship can be further quantified by considering the expected localization error:
\begin{equation}
\begin{aligned}
\mathbb{E}&[|L_r(t) - L_{gt}(t)|] \\
&= \mathbb{E} \left[|u L_\alpha(t) + (1-u) L_r(t-\Delta t) - L_{gt}(t)|\right] \\
&\leq u \mathbb{E}[|L_\alpha(t) - L_{gt}(t)|] \\
&\quad + (1-u) \mathbb{E}[|L_r(t-\Delta t) - L_{gt}(t)|] \\
&= f(\Delta t) \mathbb{E}[|L_\alpha(t) - L_{gt}(t)|] \\
&\quad + (1-f(\Delta t)) \mathbb{E}[|L_r(t-\Delta t) - L_{gt}(t)|],
\end{aligned}
\end{equation}
where $L_{gt}(t)$ is the ground truth localization at time slot $t$.

As the latency $\Delta t$ decreases, the weight $f(\Delta t)$ assigned to the DNN error term increases, and the expected localization error becomes more dependent on the accuracy of the DNN output. By minimizing the DNN inference latency, we effectively reduce the expected localization error and improve the overall accuracy of the fused localization result.

In summary, our proposed fusion method addresses the limitations of Kalman filtering for DNN-based localization by introducing a latency-aware uncertainty estimation technique. By transforming the problem of accuracy improvement into a problem of latency reduction, we enable effective fusion of DNN outputs with real-time localization results from on-vehicle sensors. The fusion equation and the latency-accuracy relationship provide a mathematical foundation for optimizing the localization performance in our system.
\section{Communication Optimization of EdgeLoc}\label{section5}

Building upon the real-time parallel architecture and localization fusion techniques, we now present the communication optimization strategies employed in EdgeLoc to ensure robust and efficient performance in dynamic network environments. The communication between vehicles and RSUs is subject to time-varying network conditions, which can significantly affect the latency of DNN inference and, consequently, the localization accuracy. 
To address this challenge, we formulate the problem of selecting the optimal DNN split point as a multi-armed bandit problem, enabling online learning to adapt to network dynamics.

\subsection{Sliding Window UCB Algorithm for Split Point Selection}
To address the exploration-exploitation trade-off in the multi-armed bandit problem and efficiently learn the optimal split point, we employ the UCB algorithm. UCB is a well-established algorithm that balances the exploration of less-chosen split points with the exploitation of the currently best-performing split point.

Consider a set of $K$ possible split points for the DNN model, denoted as $\left\{s_1, s_2, \ldots, s_K\right\}$. Each split point represents a different partition of the DNN computation between the vehicle and the RSU.
%, as described in the parallel architecture of EdgeLoc (Section 3). 
The objective is to select the split point that minimizes the expected localization error while adapting to the time-varying network conditions.
At each time step $t$, the UCB algorithm selects the split point $I_t$ based on the following rule:
$$
I_t=\underset{s_i}{\arg \max }\left(\bar{X}_i+\sqrt{\frac{16 \hat{\sigma}_{i, n_i-1}^2 \ln (t-1)}{n_i-1}}\right),
$$
where $\bar{X}_i$ is the empirical mean of the localization errors observed for the split point $s_i$ up to time slot $t-1$. In addition, $n_i$ is the number of times that the split point $s_i$ has been selected up to time $t-1$. Term $\hat{\sigma}_{i, n_i-1}^2$ is the empirical variance of the localization errors for split point $s_i$.
The empirical mean $\bar{X}_i$ is calculated as:
$$
\bar{X}_i=\frac{1}{n_i} \sum_{o=1}^{t} X_{i, o} \cdot \textbf{1}_{I_o=i},
$$
where $X_{i, o}$ represents the real localization error at the $i$-th split point and time slot $o$.

To adapt to the time-varying network conditions, we employ a sliding window variant of the UCB algorithm. Instead of considering the entire history of observations, the sliding window UCB algorithm only considers the most recent $W$ observations for each split point.
The empirical mean $\bar{X}_i$ is calculated using the observations within the sliding window:
$$
\bar{X}_i=\frac{1}{n_i(W)} \sum_{o=t-W+1}^{t} X_{i, o} \cdot \textbf{1}_{I_o=i},
$$
where $n_i(W)$ is the number of times split point $s_i$ has been selected until time slot $t-1$ within the sliding window $W$.
In particular, the sliding window UCB algorithm allows the system to quickly adapt to changes in the network conditions by focusing on the most recent observations. The window size $W$ can be tuned based on the expected rate of change in the network dynamics.
By combining the sliding window UCB algorithm with EdgeLoc's localization fusion method, the system adaptively selects the optimal split point under time-varying network conditions, minimizing the localization error. This approach ensures that the EdgeLoc system can effectively adapt to dynamic network environments and maintain high localization accuracy.

\subsection{Mitigating Network Fluctuations with Distribution Comparison}

To further enhance the robustness of the EdgeLoc system against network fluctuations, we introduce a distribution comparison mechanism in conjunction with the sliding window UCB algorithm. The system maintains Gaussian distributions of the localization errors for each split point within the sliding window.

At each time step, the system compares the current distribution of localization errors with the previous distribution for each split point. If a significant change in the distribution is detected, it indicates a potential change in the network conditions. However, to avoid triggering unnecessary retraining of the UCB algorithm due to temporary fluctuations, the system requires consistent distribution changes for a predefined number of consecutive time steps.

The sliding window size $W$ and the number of consecutive time steps required for triggering UCB retraining can be adjusted based on the specific characteristics of the network environment. For example, setting the sliding window size to 50 and requiring 3 consecutive distribution changes can provide a balance between responsiveness and stability.

By incorporating the distribution comparison mechanism and the sliding window approach, the EdgeLoc system can effectively mitigate the impact of network fluctuations and maintain robust performance in dynamic network environments.

\subsection{Seamless Integration with EdgeLoc}

The UCB-based split point selection algorithm, along with the sliding window and distribution comparison mechanisms, is seamlessly integrated into the EdgeLoc system. The system continuously monitors the network conditions, maintains the sliding window of localization errors for each split point, and compares the distributions to detect changes in the network environment.

When a consistent change in the network conditions is detected, the system triggers the retraining of the UCB algorithm to adapt to the new environment. The UCB algorithm learns the optimal split point based on the updated network conditions, ensuring optimal performance in terms of localization accuracy and inference latency.

The integration of these online learning and adaptation mechanisms into EdgeLoc enables it to dynamically adjust to varying network conditions, maintaining high localization accuracy and efficient resource utilization in infrastructure-assisted autonomous driving scenarios. By leveraging the parallel architecture (Section \ref{section3}) and the latency-accuracy relationship established in the localization fusion method (Section \ref{section4}), the communication optimization strategies presented in this section complete the EdgeLoc system, ensuring robust and adaptive performance in real-world deployments.

The mathematical proofs and detailed analysis of the adapted UCB algorithm for the EdgeLoc system are provided in the Appendix \ref{appendix}.
\section{Experiments}\label{section6}
In this section, we first introduce the experimental environment and settings. 
First, we validate the efficacy of EdgeLoc in addressing the challenges of DNN-based E2E localization and traditional sensor-based methods.
%Initially, experiments are conducted to validate the efficacy of EdgeLoc in addressing the challenges of DNN-based E2E localization and traditional sensor-based methods. 
Subsequently, we present experimental data to illustrate the impact of latency on localization performance, demonstrating the effectiveness of our proposed localization fusion method in transforming the problem of accuracy improvement into a latency reduction problem.
Finally, we perform a comprehensive system-level test in a 5G cellular network environment to assess the real-world applicability of our proposed solution, especially for different network conditions.

%evaluate the performance of 
% , targeting at assessing the system's performance and validating its effectiveness in selecting the optimal DNN split point using the adapted UCB algorithm.

% representing a potential real-world deployment scenario for the proposed EdgeLoc system. This test aims to assess the system's performance and validate its effectiveness in selecting the optimal DNN split point using the adapted UCB algorithm.

\textbf{Dataset}: Real-time localization data was from the RobotCar dataset \cite{maddern20171}, specifically utilizing VO results derived from cameras. The recorded VO datasets were streamed into the system via the ROS environment to emulate live sensor data, replicating the sensor inputs that an autonomous vehicle would provide for localization.

\textbf{Evaluation Metrics}:
The primary metric for evaluating the efficacy of EdgeLoc was localization accuracy, which served as the definitive measure of the system's performance. The system's performance was distilled into this singular metric by integrating uncertainty estimation and localization fusion techniques into EdgeLoc.

\subsection{Comparative Validation of EdgeLoc in Solving E2E and Sensor-Based Localization Problems}
\begin{figure*}
  \centering
  \includegraphics[width=1\textwidth]{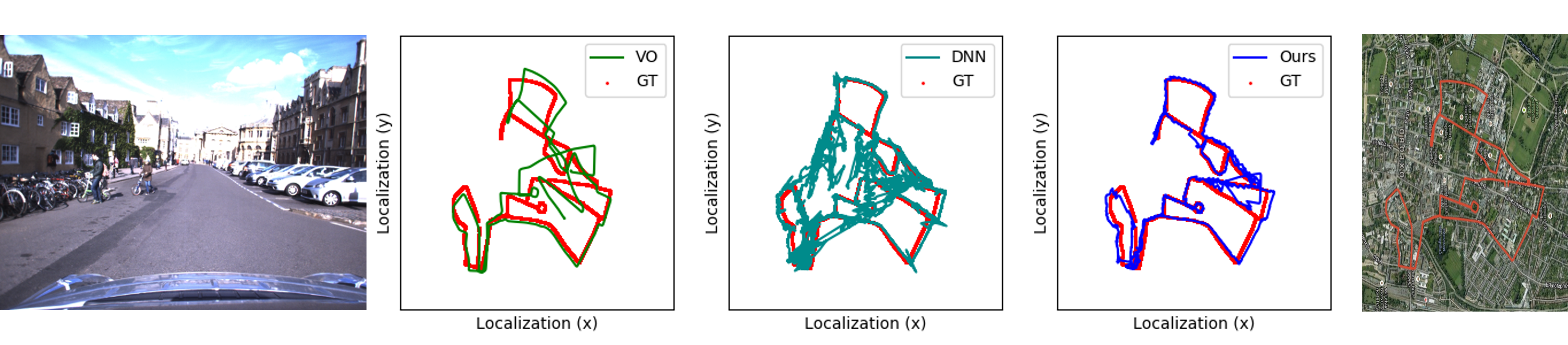}
  \caption{Comparison of localization errors for different methods over time. The methods Visual Odometry, DNN-based E2E Localization\cite{kendall2015posenet}, and EdgeLoc are respectively shown from the second to fourth images. The red curves represent the ground truth track. The left first sub-figure is an example of input image, and the right first sub-figure is the map and the real trace.}
  \label{fig:scenarioComSingle}
\end{figure*}

\begin{figure}
  \centering
  \includegraphics[width=0.4\textwidth]{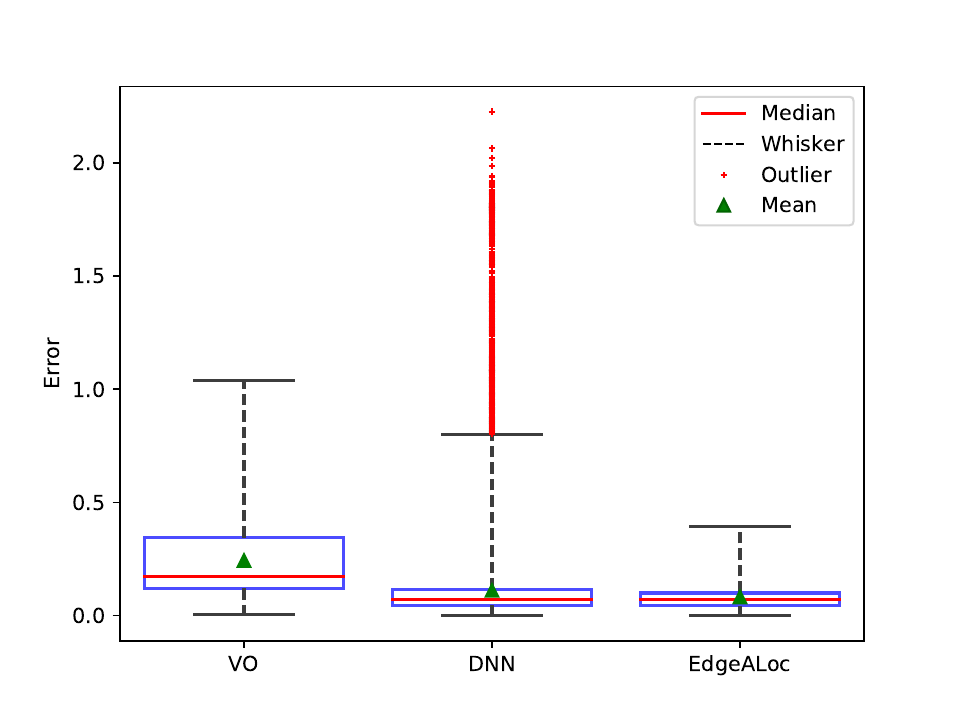}
  \caption{Localization error comparison among VO, DNN-based E2E localization, and EdgeLoc.}
  \label{fig:compare}
\end{figure}

\begin{figure}[!ht]
\centering
\subfigure[Localization results based on Kalman Filter]{
\includegraphics[width=0.18\textwidth]{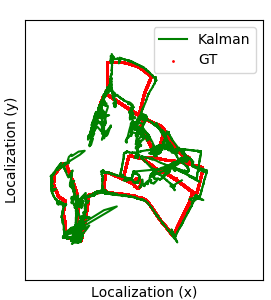}
\label{fig:subfig1}
}
\subfigure[Localization comparison between EdgeLoc, DNN, and Kalman Filter.]{
\includegraphics[width=0.28\textwidth]{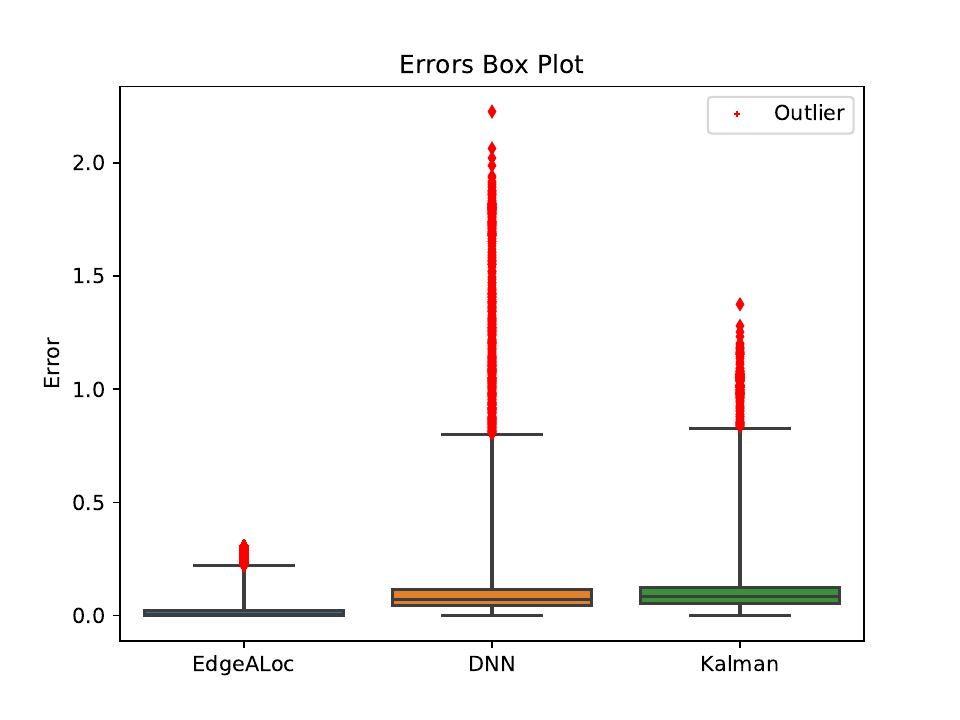}
\label{fig:subfig2}
}
\caption{Results of Comparative Experiments on Kalman Filtering}
\label{fig:klm}
\end{figure}

\begin{figure*}
  \centering
  \includegraphics[width=1\textwidth]{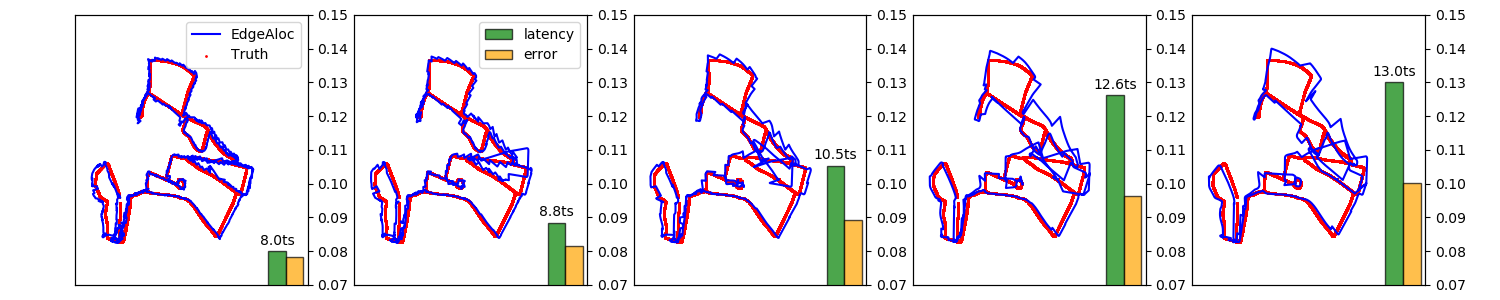}
  \caption{Impact of latency on localization performance in the EdgeLoc system. Red points indicate ground truth data for localization, while blue lines represent EdgeLoc's results. Latency is represented by green bars ("ts" here means time steps, which are used to characterize latency after normalization), and error values (Measure with y-axis value) by orange bars. The right-hand y-axis shows the scale for error values. Latency increases progressively from left to right in the subplots. Numerical mapping changes were applied for time step values.}
  \label{fig:accu2la}
\end{figure*}
To validate the effectiveness of the EdgeLoc system, we conducted experiments comparing its performance against single-method localization approaches, namely, VO and DNN based localization. Fig.~\ref{fig:scenarioComSingle} illustrates the localization errors of these three methods over time. The results reveal that the VO method suffers from cumulative errors, resulting in a gradual drift in localization accuracy. However, it is important to note that VO does not produce any outliers in its localization estimates. On the other hand, the E2E DNN approach exhibits a high degree of randomness in its localization errors, as shown in the figure. Although the majority of its localization estimates are highly accurate, the E2E DNN method generates a significant number of outliers, which can severely impact the overall localization performance.
As evident from the figure, EdgeLoc achieves consistent and accurate localization results without suffering from cumulative errors or producing outliers.

Fig.~\ref{fig:compare} provides a quantitative comparison of the localization errors for each method using box plots. The box plot corresponding to the VO method confirms that while it has the highest median error, it does not generate any outliers. In contrast, the box plot for the E2E DNN approach reveals the presence of numerous outliers, despite having a lower median error compared to VO. EdgeLoc, as shown in its corresponding box plot, not only achieves the lowest median error but also successfully eliminates outliers, demonstrating its superior performance in terms of both accuracy and reliability.

Based on the experimental results provided, we further validate the superiority of EdgeLoc over the Kalman filtering approach for localization fusion. Fig.~\ref{fig:klm} visually compares the localization results obtained using the Kalman filter and EdgeLoc. It is evident that the Kalman filter fails to effectively fuse the DNN localization outputs with the real-time sensor-based localization, resulting in significant deviations and inconsistencies in the estimated trajectory. In contrast, EdgeLoc achieves a much smoother and more accurate localization result by employing its uncertainty-aware fusion method.

Quantitatively, EdgeLoc's total localization error is \textbf{2676.35}, which is \textbf{30.26\%} lower than the Kalman filter's total error of \textbf{3837.65}. This substantial reduction highlights the limitations of the Kalman filter in handling the uncertainties and potential outliers associated with DNN localization outputs. Furthermore, EdgeLoc's total error is \textbf{29.62\%} lower than the DNN localization error of \textbf{3802.57} and\textbf{ 67.75\%} lower than the VO error of \textbf{8298.58}, demonstrating its ability to effectively leverage the strengths of both DNN and sensor-based localization methods while mitigating their individual drawbacks. 

\begin{figure}
  \centering
  \includegraphics[width=0.4\textwidth]{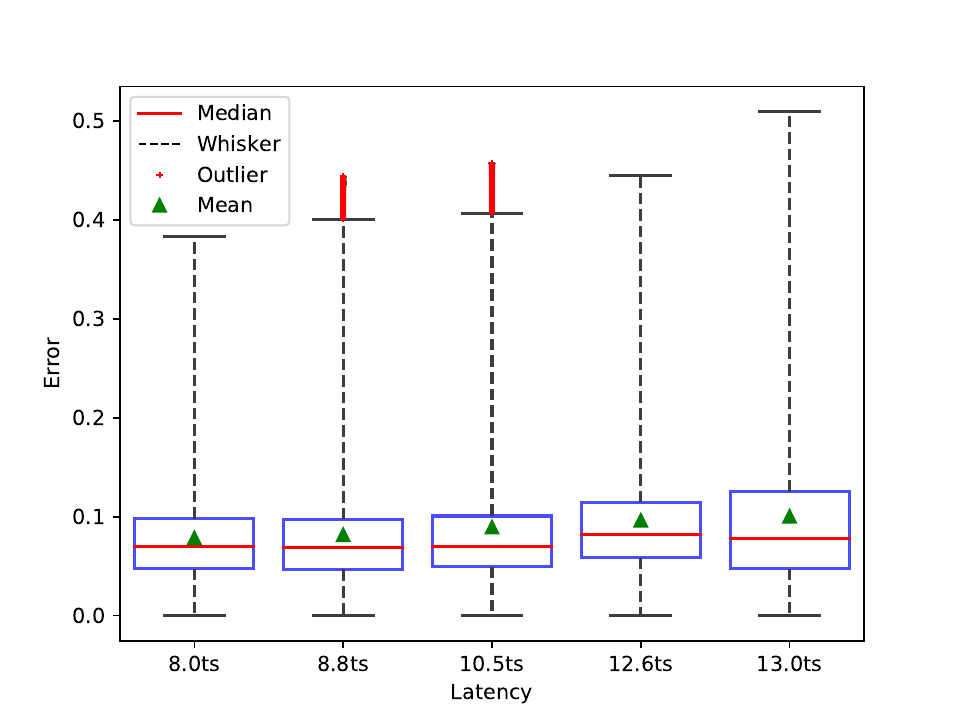}
  \caption{Box plots of localization errors for different latency ranges in the EdgeLoc system.}
  \label{fig:BoxCompare5}
\end{figure}

EdgeLoc's ability to map latency to localization accuracy is a key innovation that transforms the problem of improving accuracy into a problem of reducing latency. To showcase this relationship, we conducted experiments with varying latency conditions and evaluated their impact on localization error.

Fig.~\ref{fig:accu2la} demonstrates that as latency increases, there is a corresponding rise in error, thus validating the efficacy of the latency-to-accuracy module in the EdgeLoc framework. Fig.~\ref{fig:BoxCompare5} presents box plots of the localization errors for different latency ranges, further reinforcing the relationship between latency and accuracy. As the latency decreases, the box plots show a consistent reduction in both the median error and the spread of errors. This demonstrates that by minimizing latency, EdgeLoc effectively improves localization accuracy and reduces the variability in the localization results.

These results collectively highlight the importance of latency in the EdgeLoc system. By focusing on minimizing latency through optimized collaborative inference and efficient communication, EdgeLoc achieves improvements in localization accuracy.

\subsection{Overall System Performance in a Cellular Network Environment with Different Network Speed}
To validate the EdgeLoc system's ability to adapt to varying network conditions and consistently select the optimal DNN split point in a real-world scenario, we conducted a comprehensive experiment in a cellular network environment. This testing is crucial as it demonstrates the system's capability to effectively learn and converge to the optimal split point under realistic network conditions, where factors such as signal strength, interference, and network load may significantly impact performance.
In this experiment, we focus on evaluating EdgeLoc's performance under different network speeds, simulating various real-world scenarios. By adjusting the network bandwidth using Wondershaper, we can assess the system's ability to dynamically select the optimal split point and maintain high localization accuracy across a range of network conditions.

This section describes the experimental setup, including the 5G cellular network deployment, and discusses the results obtained, focusing on the effectiveness of the UCB algorithm for selecting the optimal DNN split point.

\subsubsection{Experimental Setup and 5G Cellular Network Deployment}

\begin{figure}
\centering
\includegraphics[width=0.49\textwidth]{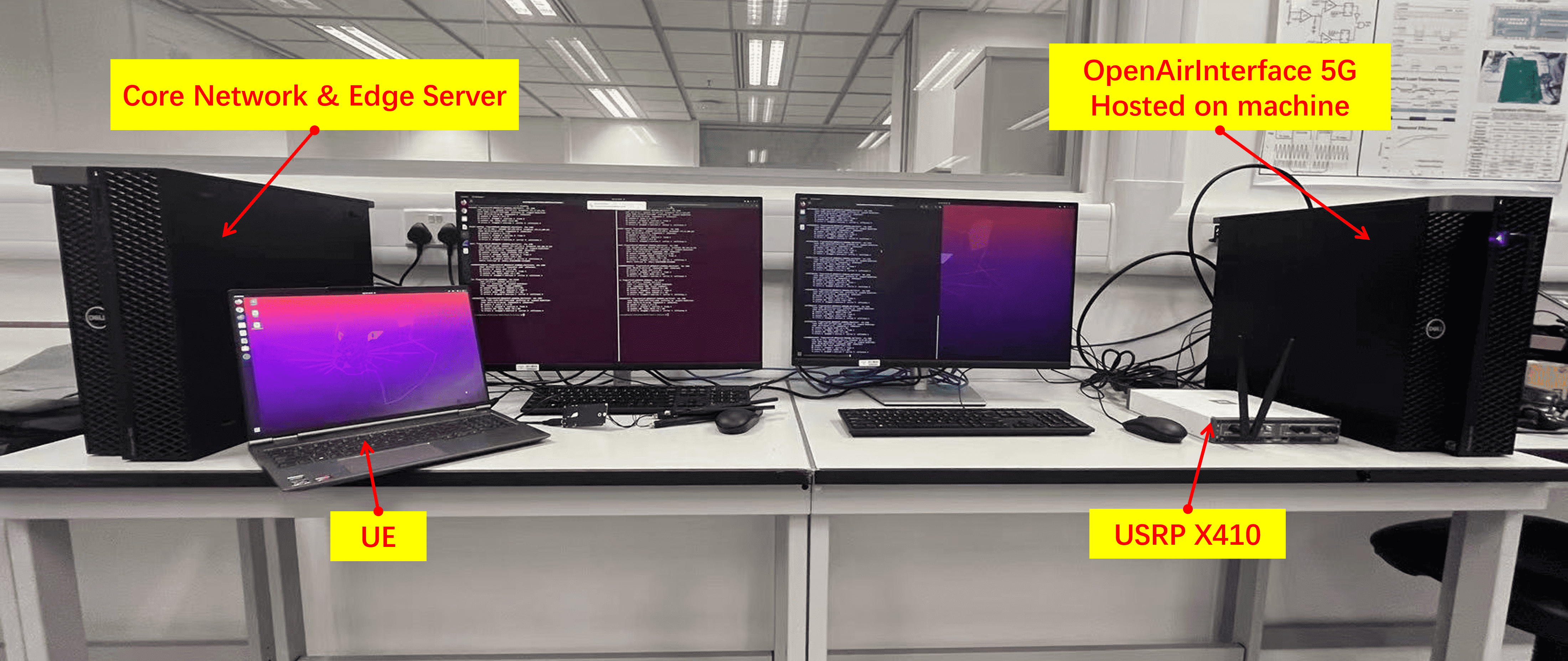}
\caption{Experimental setup for evaluating EdgeLoc's localization performance in a 5G cellular network environment.}
\label{fig:cellular_setup}
\end{figure}

Fig.~\ref{fig:cellular_setup} illustrates the experimental setup designed to evaluate EdgeLoc's performance in a 5G cellular network environment. The setup consists of three main components: User Equipment (UE), Core Network, and gNodeB (gNB). The UE represents the autonomous vehicle and is equipped with the necessary threads to execute real-time localization tasks and communicate with the gNB. The gNB acts as the base station for the cellular network, providing cellular coverage and facilitating communication between the UE and edge server in the EdgeLoc system.

The Core Network is responsible for managing communication between the UE and the gNB, ensuring reliable and secure data transmission. In this experimental setup, the Core Network machine also serves as the Edge Server, providing additional computational resources for DNN inference in the EdgeLoc system. The software system used in this experiment is OpenAirInterface \cite{nikaein2014openairinterface}, an open-source platform for 5G experimentation and prototyping. OpenAirInterface enables the deployment of a complete 5G network, including the UE, gNB, and Core Network components. We employ Wondershaper \cite{wondershaper}, a network traffic control tool, to regulate the network speed between the UE and the gNB. Control the network speed uploaded to gNB on the UE side

\subsubsection{Performance Evaluation and Results}
We conduct experiments to assess the localization accuracy achieved by EdgeLoc under different network conditions and evaluate its capability to dynamically select the optimal DNN split point.
\begin{figure}
\centering
\subfigure[The online learning process. After the star marked time, the optimal split point has been learned.]{
\includegraphics[width=0.5\textwidth]{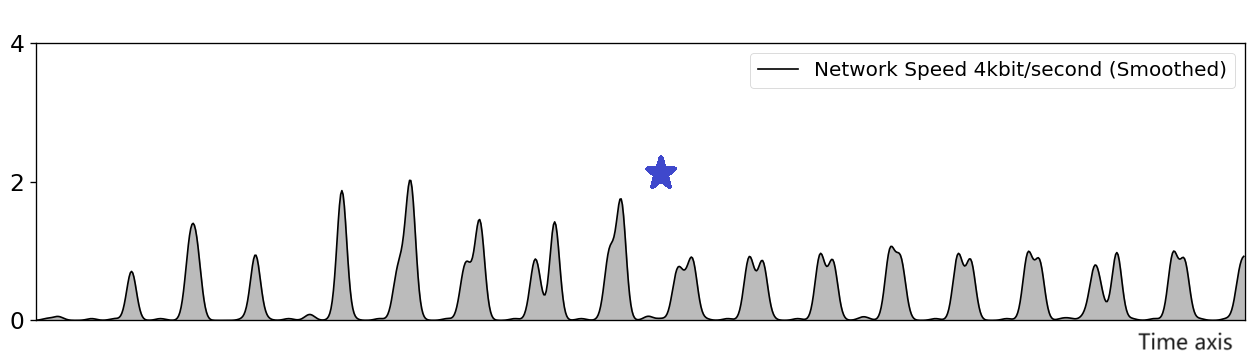}
\label{fig:subfig1}
}
\subfigure[Comparison of localization errors over time for EdgeLoc with the adapted UCB algorithm under varying network conditions.]{
\includegraphics[width=0.5\textwidth]{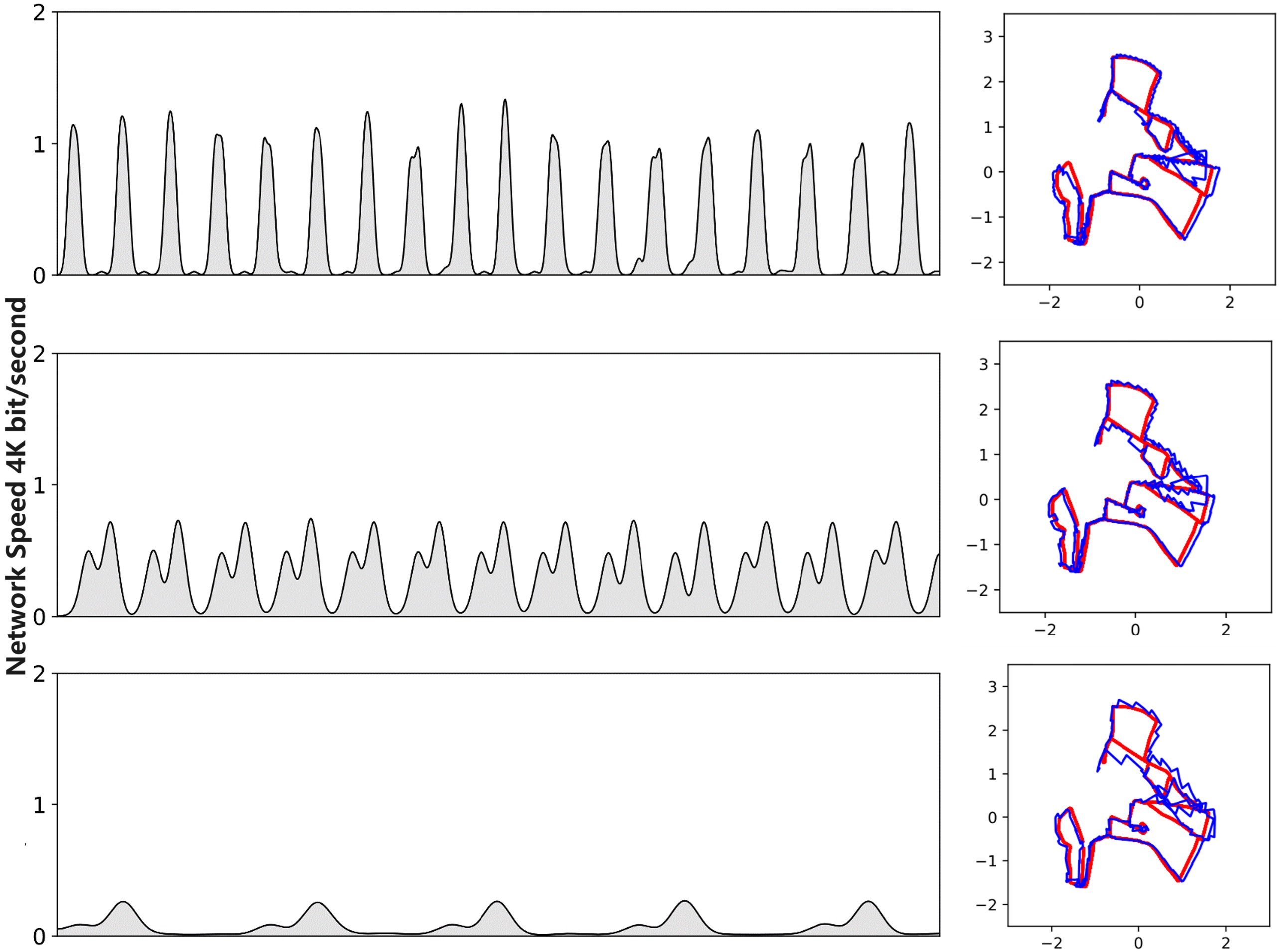}
\label{fig:subfig2}
}
\caption{Communication states of EdgeLoc in different stages and cases}
\label{fig:klm}
\end{figure}
Fig.~\ref{fig:subfig1} illustrates the online learning process of the adapted UCB algorithm in EdgeLoc. The algorithm dynamically selects different DNN split points over time, and converges to the optimal split point after a certain number of iterations (marked by the star in the figure). Fig.~\ref{fig:subfig2} illustrates the localization errors over time for EdgeLoc using the adapted UCB algorithm under three different network conditions. The experimental log results demonstrate that despite our deliberate manipulation of network speeds to create varying network environments, EdgeLoc consistently maintains its ability to select the optimal split point across the varying network scenario. By leveraging the adapted UCB algorithm to dynamically select the optimal DNN split point, EdgeLoc effectively adapts to changes in the environment and optimizes performance accordingly.
\section{conclusions}\label{section7}
This paper presents EdgeLoc, a communication-adaptive parallel system for real-time localization in infrastructure-assisted autonomous driving. EdgeLoc leverages precise localization from roadside infrastructure to enhance real-time localization based on onboard sensor-based visual odometry. The system employs a parallel processing architecture, combining the real-time performance of traditional methods with the high accuracy of deep learning approaches. EdgeLoc achieves communication adaptivity through online learning and addresses network fluctuations via window-based anomaly detection.

By auto-splitting V2I collaborative inference, EdgeLoc achieves optimal latency. Meanwhile, online distributed learning for decision-making ensures maximum localization improvement. Experimental results demonstrate that even with the most basic E2E DNN for localization estimation, our system realize a 67.75\% reduction in localization error for real-time local visual odometry, a 29.95\% reduction for non-real-time collaborative inference, and a 30.26\% reduction compared to Kalman filtering.

EdgeLoc is implemented using ROS and open-sourced for easy deployment. Future research directions include testing in larger-scale and more complex environments, as well as incorporating other types of sensors and localization algorithms. In summary, EdgeLoc provides a universal edge computing assisted solution for high-accuracy real-time localization in infrastructure-assisted autonomous driving. By leveraging the edge computing resources at RSUs and integrating them with on-vehicle systems, EdgeLoc pushes the frontier of collaborative perception and decision-making in intelligent transportation systems.
The system demonstrates the advantages of combining traditional methods with deep learning approaches, as well as the benefits of parallel computing architecture and online learning techniques. The system is open-sourced and thus has potential value in research on infrastructure-assisted autonomous driving, connected vehicles, etc.
\appendices
\section{Localization Error Analysis of ParaLoc}\label{appendix}
In the scenario of modeling errors in localization systems, this can be characterized based on the $K$-armed bandit problem. Consider random variables $E_{i,n}$ for $1 \leq i \leq K$ and $n \geq 1$, where each $i$ represents a specific action, namely a splitting point. The variable $E_{i,t}$, representing the localization error, is independent and identically distributed with respect to indices $i$ and $t$. Let $\mu_i$ denote the mean value of the random variable $E_{i,n}$ at action $i$. The optimal action's mean value is denoted by $\mu^{\ast} = \max_i \mu_i$.

A policy in the context of localization error modeling can be considered as an algorithm that selects the next splitting point to use, based on the sequence of past selections and their associated localization errors. In this scenario, the UCB1-normal policy, as described before, is a well-known algorithm for the bandit problem, particularly suitable for processes with Gaussian-distributed errors.

For convenience, we reconstruct the UCB index as \cite{Auer2002Finite}:
\begin{equation}\label{ucb}
    \Phi_{i,n} = \bar{X}_{T_i(n)} + \sqrt{ \frac{16\hat{\sigma}^2_{i,n} \ln(n-1)}{T_i(n)-1}},
\end{equation}
where
\begin{equation}
 \bar{X}_{T_i(n)} = \frac{1}{T_i(n)} \sum_{t=1}^n X_{i,t} ~1_{I_t=i},
\end{equation}
and
\begin{equation}
    \hat{\sigma}^2_{i,n} = \frac{1}{T_i(n)} \sum_{t=1}^n X^2_{i,t} 1_{I_t=i} - \bar{X}^2_{T_i(n)}.
\end{equation}
Term $T_i(n) = \sum_{t=1}^n 1_{I_t =i}$ denotes the number of times the player selected arm $i$ until time slot $n$. 
%Here, $T_i$ represents the number of times the splitting point $i$ has been used thus far. 
This approach aims to minimize the localization error by dynamically choosing the splitting point with the highest expected reduction in error, as quantified by the UCB index. 
The localization system is as follows:
\begin{equation}\label{localization}
\begin{cases}
L_r(t) = u L_\alpha(t) + (1-u) L_r(t-\Delta t), \\ 
\qquad \text{if } t = x \Delta t, x \in \mathbb{N}, \\
L_r(t+a) = L_r(t+a-1) + L_\beta(t+a) - L_\beta(t+a-1), \\
\qquad \text{if } t \neq x \Delta t, x \in \mathbb{N}; a \in [1, \Delta t]. \\
\end{cases}
\end{equation}

\noindent \textbf{Theorem:}\quad
For the $K$-armed bandit problem with Gaussian reward process, the localization upper bound after $n$ plays by running the UCB1-nromal policy is ${R}_n \leq f(L_\alpha, L_\beta)$.
\noindent \textbf{Proof:}\quad We will provide a proof by combining the UCB algorithm and the mathematical model of the proposed localization system. We firstly define:
\begin{equation}
    X_{i, t} = L(t) - L_r(i, t),
\end{equation}
where $X$ represents localization errors, and $i$ represents the $i^{th}$ action (split point). $X_{i, t}$ represents the practical localization error at the $i^{th}$ split point that was selected at time step $t$. Based on \eqref{localization}, we can get:
\begin{equation}\label{error}
\begin{split}
    & \sigma^2_{n} = \frac{1}{n}\sum_{t=1}^n (L(t) - L_r(t))^2 \\
    & - \left( \frac{1}{t} \left( \int_0^t \left[ L(t) - (u \times L_\alpha(t) + (1 - u) L_r(t)) \right] dt \right. \right. \\
    & + \left. \left. \int_0^t \left[ L(t) - L(t-1) \right] dt - \int_0^t \left[ L(t) - L_\beta(t-1) \right] dt \right) \right)^2,
\end{split}
\end{equation}
where
\begin{equation}\label{uncertainty}
    u =\int_0^{\Delta t}(L(t)-L(t-\Delta t))-\left(L_\beta(t)-L_\beta(t-\Delta t)\right).
\end{equation}

At each time slot $t$, a system pulls an action $I_t$ according to the UCB1-normal policy and receives a reward (localization error) $X_{I_t,t}$ from a Gaussian distribution with unknown mean and variance.
After $n$ rounds, the accumulated regret is defined as
\begin{equation}\label{regret01}
    R_n = \max_{i=1,\ldots, K} \sum_{t=1}^{n} X_{i,t} - \sum_{t=1}^n X_{I_t,t}.
\end{equation}
%According to:
%\begin{equation}\label{mean}
% \bar{E}_i = \frac{1}{n} \sum_{t=1}^n E_{i,t},
%\end{equation}
The pseudo-regret is
\begin{equation}\label{regret02}
\begin{split}
    \bar{R}_n &= \max_{i=1,\ldots, K} \mathbb{E} \left[ \sum_{t=1}^{n} X_{i,t} - \sum_{t=1}^n X_{I_t,t} \right]
    =n \mu^{\ast} - \sum_{t=1}^n \mathbb{E} [\mu_{I_t}].
\end{split}
\end{equation}
Let $\Delta_i = \mu^{\ast} - \mu$ be the suboptimality parameter of arm $i$.
Then, the pseudo-regret can be rewritten as
\begin{equation}\label{regret02}
\begin{split}
    \bar{R}_n & = \left(\sum_{i=1}^K \mathbb{E} [T_i(n)] \right) \mu^{\ast} - \mathbb{E} \left[\sum_{i=1}^K T_i(n)\mu_i\right], \\
    & = \sum_{i=1}^K \Delta_i \mathbb{E} [T_i(n)].
\end{split}
\end{equation}

According to Eq. \eqref{regret02}, we can bound the regret $\bar{R}_n$ by simply bounding each $\mathbb{E} [T_i(n)]$.
For each $t>1$, we bound the probability of the indicator function that $I_t = i$. 
Specifically, let $L$ be an arbitrary positive integer and
\begin{equation}
    c_{t,s} = \sqrt{ \frac{16\hat{\sigma}^2_{i,s} \ln(t-1)}{s_i-1}}. 
\end{equation}
Then, we have
\begin{equation}\label{MainProof}
    \begin{split}
        T_i(n) &=1 + \sum_{t=K}^{n} \left\{I_t = i \right\},\\
        &\leq L + \sum_{t=K}^{n} \left\{ I_t = i, T_i(t-1) \geq L \right\}, \\
        &\leq L + \sum_{t=K}^{n} \left\{ \bar{X}_{T^{\ast}(t-1)} + c_{t-1, T^{\ast}(t-1)} \leq \bar{X}_{T_{i}(t-1)}\right. \\
        &\left. \qquad  + c_{t-1, T_i(t-1)}, T_i(t-1) \geq L  \right\}, \\
        &\leq L + \sum_{t=K}^{n} \left\{\min_{0<s<t} \bar{X}^{\ast}_s + c_{t-1, s} \leq \max_{L<s_i<t} \bar{X}_{i,s_i} + c_{t-1,s_i} \right\},\\
        &\leq L + \sum_{t=1}^{\infty}\sum_{s=1}^{t-1}\sum_{s_i=L}^{t-1} \left\{\bar{X}^{\ast}_s + c_{t, s} \leq  \bar{X}_{i,s_i} + c_{t,s_i} \right\}.\\
    \end{split}
\end{equation}
Note that $\bar{X}^{\ast}_s + c_{t, s} \leq  \bar{X}_{i,s_i} + c_{t,s_i}$ indicates that at least one of the three events must hold: 
(i) $\bar{X}^{\ast}_s \leq \mu^{\ast} - c_{t,s} $; (ii) $\bar{X}_{i,s_i}\geq \mu_i + c_{t, s_i}$;  (iii) $\mu^{\ast} \leq \mu_i + 2c_{t, s_i} $.
Thus, Eq. \eqref{MainProof} can be rewritten as
\begin{equation}\label{MainProof01}
\begin{split}
     T_i(n)\leq& L + \sum_{t=1}^{\infty}\sum_{s=1}^{t-1}\sum_{s_i=L}^{t-1} \left\{ \mathrm{Pr}\left\{\bar{X}^{\ast}_s \leq \mu^{\ast} - c_{t,s}\right\} \right.\\
     &\left. +\mathrm{Pr}\left\{\bar{X}_{i,s_i}\geq \mu_i + c_{t, s_i} \right\} +\mathrm{Pr}\left\{\mu^{\ast} \leq \mu_i + 2c_{t, s_i} \right\} \right\}.\\
\end{split}
\end{equation}

Next, we bound the probabilities of events (i), (ii), and (iii).
The random variable $(\bar{X}_{i,s_i} - \mu_i)/\sqrt{\hat{\sigma}^2_{i,s_i}/(s_i-1)}$ has a student distribution with $s_i-1$ degrees of freedom.
Then, using the fact that $\mathrm{Pr} \left\{ X\geq a \right\} \leq e^{-a^2/4}$ for the student distribution, we have
\begin{equation}\label{even01}
\begin{split}
    \mathrm{Pr}\left\{\bar{X}^{\ast}_s \leq \mu^{\ast} - c_{t,s}\right\} 
    &=\mathrm{Pr}\left\{ \frac{\mu^{\ast} - \bar{X}^{\ast}_s} {\sqrt{\hat{\sigma}^2_{\ast,s}/(s-1)}} \geq 4\sqrt{\ln t}\right\},\\
   &\leq t^{-4},
\end{split}
\end{equation}
and
\begin{equation}\label{even02}
\begin{split}
\mathrm{Pr}\left( \bar{X}_{i,s_i} \geq \mu_i + c_{t, s_i} \right)
&= \mathrm{Pr}\left( \frac{\bar{X}_{i,s_i} - \mu_i}{\sqrt{\hat{\sigma}^{i, s_i}/(s_i-1)}} \geq 4\sqrt{\ln t}\right), \\
&\leq t^{-4},
\end{split}
\end{equation}
for all $s_i\geq 8\ln t$ and $a = 4\sqrt{\ln t}$.
For event (iii), the random variable $s_i \hat{\sigma}^2_{t, s_i}/\sigma^2_{i}$ has the $\chi^2$-distributed with $s_i$ degrees of freedom \cite{Wilks1962}.
Thus, by using the fact that $\mathrm{Pr} \left\{ X\geq 4s \right\} \leq e^{-(s+1)/2}$ with $s = s_i-1$, we obtain
\begin{equation}\label{even03}
\begin{split}
    &\mathrm{Pr}\left\{\mu^{\ast} \leq \mu_i + 2c_{t, s_i} \right\}, \\
    & =\mathrm{Pr}\left\{ \frac{s_i \hat{\sigma}^2_{i, s_i}} {\sigma^2_{i}} > \frac{(s_i-1)\Delta^2_i s_i}{64 \sigma^2_i \ln t}       \right\},\\
   & \leq \mathrm{Pr}\left\{ \frac{s_i \hat{\sigma}^2_{i, s_i}}{\sigma^2_{i}} > 4(s_i -1) \right\},\\
   &\leq e^{-s_i/2} \leq t^{-4},
\end{split}
\end{equation}
for $s_i\geq \max\{256\sigma^2_i/\Delta^2_i, 8\}\ln t$. 
Thus, we can set $L = \max\{256\sigma^2_i/\Delta^2_i, 8\}\ln t$.
By substituting Eqs. \eqref{even01}, \eqref{even02}, and \eqref{even03} into Eq. \eqref{MainProof01}, it yields
\begin{equation}\label{MainProof02}
\begin{split}
     \mathbb{E}[T_i(n)]&\leq \frac{8\ln n}{\Delta_i^2} + \sum_{t=1}^{\infty}\sum_{s=1}^{t-1}\sum_{s_i=L}^{t-1} 3t^{-4}  \\
     &\leq \frac{256 \sigma^2_i \ln n }{\Delta_i^2} + 8 \ln n + \frac{\pi^4}{30}.\\
\end{split}
\end{equation}
Therefore, the regret upper bound can be computed as 
\begin{equation}\label{regret03}
\begin{split}
    \bar{R}_n &= \sum_{i=1}^K \Delta_i \mathbb{E} [T_i(n)]\\
    & \leq 256 \ln n \sum_{i, i\neq i^{\ast}} \frac{\sigma^2_i}{\Delta_i} + (8 \ln n + \frac{\pi^4}{30})\sum_{j=1}^{K} 
     \Delta_i.
\end{split}
\end{equation}
From the above formula, it can be seen that the regret upper bound is only related to \(\sigma\) and \(\Delta\), where \(\Delta\) is an inherent property. Regarding \(\sigma\), based on \eqref{error}, we have
\begin{equation}
\begin{split}    
    \sigma^2_{n} =& \frac{1}{n}\sum_{t=1}^n (L(t) - Lr(t))^2 \\
    &-  \frac{1}{t} \left( \int_0^t \left[ L(t) - (u \times L_\alpha(t) + (1 - u) L r(t)) \right] \right. \\
    &+ \left. \int_0^t \left[ L(t) - L(t-1) \right] - \left[ L(t) - L_\beta(t-1) \right] \right) ^2.
\end{split}
\end{equation}
We see that \(\sigma\) is only related to \(L_\alpha\), \(L_\beta\), and \(u\). According to \eqref{uncertainty}, \(u\) is solely dependent on \(\Delta t\), which corresponds to each split point and thus is an inherent property. Therefore, \(\sigma\) is only related to \(L_\alpha\) and \(L_\beta\). Consequently, we can conclude that the upper bound of \(\bar{R}_n\) \text{is only related with} \(L_\alpha\) and \(L_\beta\).

To sum up, the regret upper bound provides a theoretical foundation for the optimality of the ParaLoc system. By proving that the regret upper bound of the UCB1-normal policy only depends on the end-to-end DNN localization results ($L_\alpha$) and the real-time sensor-based localization results ($L_\beta$), we demonstrate that the ParaLoc system fully utilizes the information provided by these two variables. This suggests that, as $L_\alpha$ and $L_\beta$ are intrinsic components, no additional enhancements can be made to the theoretical worst-case performance.

% trigger a \newpage just before the given reference
% number - used to balance the columns on the last page
% adjust value as needed - may need to be readjusted if
% the document is modified later
%\IEEEtriggeratref{8}
% The "triggered" command can be changed if desired:
%\IEEEtriggercmd{\enlargethispage{-5in}}

% references section

% can use a bibliography generated by BibTeX as a .bbl file
% BibTeX documentation can be easily obtained at:
% http://mirror.ctan.org/biblio/bibtex/contrib/doc/
% The IEEEtran BibTeX style support page is at:
% http://www.michaelshell.org/tex/ieeetran/bibtex/
%\bibliographystyle{IEEEtran}
% argument is your BibTeX string definitions and bibliography database(s)
%\bibliography{IEEEabrv,../bib/paper}
%
% <OR> manually copy in the resultant .bbl file
% set second argument of \begin to the number of references
% (used to reserve space for the reference number labels box)

\bibliographystyle{ieeetr}
\bibliography{references}

\ifCLASSOPTIONcaptionsoff
  \newpage
\fi
\end{document}